\documentclass[10pt]{article}
\pdfoutput=1
\usepackage{ifpdf}

\usepackage[a4paper,top=3.0cm,width=16cm, left=2.6cm]{geometry}
\usepackage{graphicx}
\usepackage{microtype}
\usepackage[utf8x]{inputenc}
\usepackage[T1]{fontenc}
\usepackage{ae}
\usepackage{aecompl}
\usepackage{amsmath}
\usepackage{amssymb}
\usepackage{mathdots}

\usepackage[usenames,dvipsnames]{xcolor}

\usepackage[colorlinks,hyperindex, 
            bookmarks=true,bookmarksopen=true]{hyperref}
    \hypersetup
    {
        colorlinks=true,
        linkcolor=MidnightBlue,
        urlcolor=MidnightBlue,
        filecolor=black,
        citecolor=RedOrange,
        pdfstartview=FitV,
        pdftitle={},
        pdfauthor={Ilmar Gahramanov, Deniz Bozkurt},
        pdfsubject={},
        pdfkeywords={},
        pdfpagemode=None,
        bookmarksopen=true
    }

\usepackage[small,sf,bf,hang]{caption}
\usepackage[english]{babel}
\usepackage{datetime} 
\shortdate 

\usepackage{authblk}

\title{{\bf Pentagon identities arising\\ in supersymmetric gauge theory computations}}
\author[1]{Deniz N. Bozkurt}
\author[2,3,4]{Ilmar Gahramanov}
\affil[1]{Department of Physics, Koc University, 34450 Sariyer,Istanbul, Turkey}
\affil[2]{Physics Department, Mimar Sinan Fine Arts University, Bomonti 34380, Istanbul, Turkey}
\affil[3]{Department of Mathematics, Khazar University, Mehseti St. 41, AZ1096, Baku, Azerbaijan}
\affil[4]{Max Planck Institute for Gravitational Physics, Am M\"{u}hlenberg 1, D-14476 Potsdam, Germany}
\date{}

\makeatother

\begin{document}

\maketitle

\thispagestyle{empty}
 
\hfill {\textbf{Dedicated to the memory of Ludvig D. Faddeev     ~~~~~~~~}}
\vskip 0.3cm

\begin{abstract} The partition functions of three-dimensional $\mathcal N=2$ supersymmetric gauge theories on different manifolds can be expressed as q-hypergeometric integrals. By comparing the partition functions of three-dimensional mirror dual theories, one finds complicated integral identities. In some cases, these identities can be written in the form of pentagon relations. Such identities often have an interpretation as the Pachner's 3-2 move for triangulated manifolds via the so-called 3d-3d correspondence. From the physics perspective, another important application of pentagon identities is that they may be used to construct new solutions to the quantum Yang-Baxter equation.
\end{abstract}


\section{Introduction}

In this article, we  consider some pentagon relations arising from supersymmetric gauge theory computations. Pentagon identities appear in many branches of modern mathematical physics, such that exactly solvable models, two-dimensional conformal field theory, Hopf algebras,  topological field theories, knot invariants etc (see e.g. \cite{allman2017quantum,dimakis2015simplex, Gahramanov:2014ona} and references therein).

The computation of partition function of supersymmetric gauge theories on compact manifolds can be reduced to matrix integrals by using the supersymmetric localization technique \cite{Pestun:2007rz}. Such matrix integrals in case of three-dimensional supersymmetric gauge theories can be expressed in terms of q-hypergeometric integrals and/or sums. A key point is that by studying partition functions of supersymmetric dual theories one may obtain new complicated identities for this type of special functions \cite{Spiridonov:2009za,Spiridonov:2011hf,Gahramanov:2015tta,Gahramanov:2016wxi,ArabiArdehali:2017fsp}. In the paper, we consider a special type of such identities, namely five-term relations or the so-called pentagon identities.

A typical example of the pentagon identity arising from the supersymmetric gauge theories is the equality of the partition functions of three-dimensional $\mathcal N=2$ mirror dual theories which has the following form \cite{Kashaev:2012cz,Dimofte:2011ju,Dimofte:2011py,Gahramanov:2013rda,Gahramanov:2014ona,Gahramanov:2016wxi,Imamura:2012rq}
\begin{equation} \label{genpent}
\oint d\mu \; B_c \; B_c = B_c \; B_c \; B_c \; ,
\end{equation}
where the integral is over the $U(1)$ gauge fugacity and $B_c$ is the contribution of the chiral multiplet (or combination of such contributions).

Here we present some examples of pentagon relations relevant to $\mathbb{S}^3_b$, $\mathbb{S}^2 \times \mathbb{S}^1$, $\mathbb{RP}^2 \times \mathbb{S}^1$ supersymmetric partition functions.

In the context of supersymmetric gauge theories, five-term relations are interesting from the following point of view. There is a recently proposed relation called 3d-3d correspondence \cite{Dimofte:2011ju,Dimofte:2011py} (see also \cite{Galakhov:2012hy,Dimofte:2014ija,Terashima:2011xe}) in a similar spirit of the AGT correspondence \cite{Alday:2009aq}. The general idea is that one can associate a 3-manifold $M_3$ with a three-dimensional $\mathcal N = 2$ supersymmetric gauge theory denoted by $T[M_3]$  obtained from the twisted compactification of the $6d$ $\mathcal N=(2,0)$ theory on the three-manifold $M_3$. This correspondence translates the ideal triangulation of the 3-manifold into mirror symmetry for three-dimensional supersymmetric theories. Independence of the corresponding 3-manifold invariant on the choice of triangulation corresponds to the equality of partition functions of mirror dual theories. In this context the identity (\ref{genpent}) encodes a 3–2 Pachner move for 3-manifolds.


In this article, we review some pentagon identities inspired by supersymmetric gauge theory computations and present a few new results. 
It is worth to mention that the importance of pentagon identities\footnote{Note that in the Faddeev's works the pentagon identity refers to the operator equation, rather than integral relations we have here.} in mathematical physics, mainly in the context of integrable models, quantum groups and knot theory has been first noticed by Ludvig Faddeev \cite{Faddeev:1993rs,Faddeev:2012zu} and his coworkers \cite{Volkov:1996ex,Volkov:2011vk,Kashaev:1996rz,Kashaev:2000ku,Kashaev:2014rea}. 

This paper is organized as follows. In section 2 we review the localization technique which enables us to compute partition functions of supersymmetric theories exactly. In section 3 we recall the three-dimensional mirror symmetry. In section 4 we present pentagon identities. We conclude with a few remarks and open problems in the last section. 

\section{Localization}

In this section, we review the basic idea of supersymmetric localization\footnote{Localization technique has a long history in topological field theory, and particular example is an instanton counting by using omega deformation \cite{Nekrasov:2002qd}.}. More details on the subject can be found in the original paper of Pestun \cite{Pestun:2007rz} and in the review papers \cite{Hosomichi:2015jta,Willett:2016adv,Cremonesi:2014dva}.

In this paper we will use partition functions
\begin{equation} \label{eucpart}
Z_{M_3} \ = \ \int D\phi \; e^{-S[\phi]} \; ,
\end{equation}
of supersymmetric gauge theories on compact manifold\footnote{In order to make the path integral well-defined one considers compact space which provides infrared cut-off.}  $M_3$, where $\phi$ denotes all the fields in a theory.  As a compact manifold we will consider $\mathbb{S}_b^3$, $\mathbb{S}^2 \times \mathbb{S}^1$ and $\mathbb{RP}^2 \times \mathbb{S}^1$.  The so-called supersymetric localization technique \cite{Pestun:2007rz} allows us to compute the partition function (\ref{eucpart}) on such manifolds exactly. The main idea of supersymmetric localization is the following\footnote{Note that in three dimensions the supersymmetric localization is technically simpler since there are no instanton corrections. Even sum over monopole charges in case of $S^2 \times S^1$ is simpler than instantons on $S^4$}.

Suppose we consider a supersymmetric gauge theory on a manifold $M$ with a fermionic operator\footnote{It is some linear combination of supercharge} $Q$. From supersymmetric invariance we have
\begin{equation} \label{QS}
QS \ = \ 0 \;.
\end{equation}
Then we can rewrite the path-integral in the following way
\begin{equation}
Z(t) \ = \ \int D \phi e^{-S[\phi]-tQ V[\phi]} \; ,
\end{equation}
with $\delta_Q V=0$, where $V$ is some functional of fields and $\delta_Q$ is a supersymmetric transformation satisfying (\ref{QS}). If the measure is $Q$-invariant\footnote{The supersymmetry is not broken in the vacuum.} one can show that the partition function $Z$ is independent of parameter $t$ and for large $t$ the path integral only gets contributions near $\delta_Q V[\phi_0]=0$
\begin{equation}
Z \ = \ \int D \phi_0 \; e^{-S[\phi_0]} \; Z_{{1-loop}}[\phi_0] \;. 
\end{equation}

\section{Mirror symmetry}

In this section, we discuss mirror symmetry in three-dimensional $\mathcal N = 2$ supersymmetric gauge theories, which will be our main focus in obtaining pentagon identities.

Three dimensional mirror symmetry was first introduced for $\mathcal N = 4$ supersymmetric gauge theories in \cite{Intriligator:1996ex} and   was extended to $\mathcal N=2$ gauge theories \cite{Aharony:1997bx}. The simplest example of $\mathcal N = 2$ mirror symmetry is the duality between supersymmetric quantum electrodynamics with one flavor and the free Wess-Zumino theory \cite{Intriligator:1996ex, Aharony:1997bx}, which are defined in the UV region and flow to the same IR fixed point:

\begin{itemize}
\item The $\mathcal N= 2$ supersymmetric quantum electrodynamics has one flavor consisting of two chiral fields $Q$, $\tilde{Q}$ and one vector multiplet $V$. This theory possesses extra U$(1)$ global symmetries: one is the topological U$(1)_{J}$, and the other is the flavor symmetry U$(1)_{A}$. 
		\begin{center}
			\begin{tabular}{|c|c|c|c|} \hline
			& U$(1)$ & U$(1)_{J}$ & U$(1)_{A}$  \\ \hline
			$Q$ & $1$ & $0$ & $1$ \\ \hline
			$\tilde{Q}$ & $- 1$ & $0$ & $1$  \\ \hline
			\end{tabular}
			\\ \vspace{0.4cm} 
			\textbf{Charges in the SQED.}
		\end{center}

\item  The mirror theory, free Wess-Zumino model\footnote{This theory is often is called the XYZ model in the literature.} is the theory containing three chiral fields $q, \tilde{q}$, and $S$ interacting through the trilinear superpotential $W = \tilde{q} S q$. This theory has two U$(1)$ global symmetries, named U$(1)_{V}$ and U$(1)_{A}$ \cite{Kapustin:2011jm}.
\begin{center}
			\begin{tabular}{|c|c|c|} \hline
			& U$(1)_{V}$ & U$(1)_{A}$ \\ \hline 
			$q$ & $1$ & $-1$  \\ \hline
			$\tilde{q}$ & $- 1$ & $-1$  \\ \hline
			$S$ & 0 & $2$  \\ \hline
			\end{tabular}
				\\ \vspace{0.4cm} 
			\textbf{Charges in the free Wess-Zumino theory.}
		\end{center}

\end{itemize}

In the context of mirror symmetry, we can identify U$(1)_{J}$ and U$(1)_{A}$ of the supersymmetric quantum electrodynamics with U$(1)_{V}$ and U$(1)_{A}$ of the Wess-Zumino model, respectively. 

\section{Pentagon identities}

Since theories discussed in the previous chapter are mirror dual to each other, the supersymmetric partition functions should agree. The partition function of the supersymmetric quantum electrodynamics discussed above has contributions of two quarks and the partition function of the mirror partner, the Wess-Zumino theory contains contributions of one meson and two singlets.

Let us first consider the three-dimensional squashed-sphere $S_b^3$ partition function\footnote{A three dimensional sphere partition function was first studied in \cite{Kapustin:2009kz}
for $\mathcal N > 2$. The extension to $\mathcal N = 2$ was done in \cite{Hama:2010av,Jafferis:2010un} for a round sphere and in \cite{Hama:2011ea} for a squashed sphere.}. The  partition function on the squashed sphere can be written in the form of hyperbolic hypergeometric functions. The mirror duality implies the following pentagon identity\footnote{Note that the equality of partition functions holds true for arbitrary R-charge of
the quarks in the IR fixed point, therefore one does not need to specify it.} \cite{Dimofte:2011ju}
\begin{align} \label{firstpent}
 \int dz s_b(y-z) s_b(y+z)  \ = \ s_b(2 y - \tfrac{i Q}{2}) \; s_b(\tfrac{i Q}{2}-y) \; s_b(\tfrac{i Q}{2} -y) \; ,
\end{align}
where the left hand-side represents the partition function of supersymmetric QED and the right side is the partition function of the dual theory. Here the double-sine function\footnote{The double sine function is a variant of Faddeev’s non-compact quantum dilogarithm \cite{Faddeev:1993rs}. There are different notations and modifications of this function, relations between some of them can be found in \cite{Spiridonov:2011hf,Gahramanov:2016ilb,Gahramanov:2017ysd}.} is defined as
\begin{equation}
s_b(x) = e^{-\frac{i\pi}{2}x^2}\prod_{j=1}^\infty\frac{1+e^{2\pi bx+2\pi i b^2(j-\tfrac12)}}
 {1+e^{2\pi b^{-1}x+2\pi ib^{-2}(\tfrac12-j)}}.
\end{equation}
The pentagon identity (\ref{firstpent}) is well-known in the literature and appeared in different subjects, mainly in the Liouville theory.

Next we consider the $S^2 \times S^1$ partition functions. The partition function can be evaluated by the localization technique, leading to the matrix integral in terms of basic hypergeometric functions. Mirror symmetry leads to the following identity for the partition functions\footnote{This identity first was proven for the case $m=0$ \cite{Kapustin:2011jm}.}:
\begin{align} \label{secondpent}
& \sum_{s \in \mathbb{Z}} \oint \frac{dz}{2 \pi i z} (-w)^s z^{n-s}\frac{(z \alpha^{-1} q^{\frac{m+ s}{2}+\frac34};q)_\infty}{(z^{-1} \alpha q^{\frac{m+s}{2}+\frac14};q)_\infty}\frac{(z \alpha^{-1} q^{\frac{m-s}{2}+\frac34};q)_\infty}{(z \alpha q^{\frac{m- s}{2}+\frac14};q)_\infty} \\ \nonumber
 & \qquad \qquad  \qquad = (-w)^n\frac{(\alpha w q^{\frac{m + n}{2}+\frac34};q)_\infty}{(\alpha^{-1} w^{-1} q^{\frac{m - n}{2}+\frac14};q)_\infty} \frac{(\alpha w q^{\frac{m - n}{2}+\frac34};q)_{\infty}(\alpha^{-2} q^{m+\frac12};q)_\infty}{(\alpha^{-1} w q^{\frac{m + n}{2}+\frac14};q)_{\infty}(\alpha^2 q^{{m}+\frac12};q)_\infty}\;,
\end{align}
where $\alpha$ and $m$ denote fugacity and the monopole charge for the axial $U(1)_A$ symmetry, $\omega$ and $n$ denote the fugacity and monopole charge for the topological $U(1)_J$ symmetry, respectively. The fugacity $z$ and the discrete parameter $s$ stand for the magnetic charge corresponding to the $U(1)$ gauge group. 

Let $\mathcal{B}(m;q,z)$ be the function, such that
\begin{equation}
\mathcal{B}(m,z)=\frac{(zq^{\frac{m}{2}+\frac12};q)_{\infty}}{(z^{-1}q^{\frac{m}{2}};q)_{\infty}} \; .
\end{equation}
Then the integral identity (\ref{secondpent}) can be written as a pentagon identity \cite{Krattenthaler:2011da,Gahramanov:2013rda,Kapustin:2011jm,Gahramanov:2016wxi}
\begin{align}
\sum_{s \in \mathbb{Z}} \oint \frac{dz}{2 \pi i z} (-w)^s z^{n-s}\mathcal{B}(m-s,z\alpha^{-1}q^{\frac14}) \; \mathcal{B}(m+s,z^{-1}\alpha^{-1}q^{\frac14}) \\ \nonumber
\qquad =(-w)^n\mathcal{B}(m+n,w\alpha q^{\frac14}) \; \mathcal{B}(m-n,w^{-1}\alpha q^{\frac14}) \; \mathcal{B}(m+\frac12,\alpha^{-2}q^{\frac14}) \; .
\end{align}

Another simple but more interesting example of the pentagon identity is provided by the equality of $\mathbb{RP}^2 \times \mathbb{S}^1$ partition functions. According to the mirror symmetry we have the following integral identity \cite{Tanaka:2014oda,Tanaka:2015pwa,Mori:2015urc}

\begin{align} \nonumber
q^{1/8}\frac{(q^2;q^2)_{\infty}}{(q;q^2)_{\infty}}\oint_{C_0} \,\frac{dz}{2{\pi}iz}z^s\sum^{1}_{m=0}a^{-1/2+m}q^{-1/4+m/2}\frac{(z^{-1}aq^{1+m} ;q^2)_{\infty}(zaq^{1+m} ;q^2)_{\infty}}{(za^{-1}q^{m} ;q^2)_{\infty}(z^{-1}a^{-1}q^{m} ;q^2)_{\infty}}
\\ 
= q^{-\frac{1}{8}} 
a^{-1/2-| \tilde{s} |}
\frac{( a^{-1} q^{1/2 + |\tilde{s}|} ; q^2 )_{\infty}}{( a q^{1/2+|\tilde{s}|} ; q^2 )_{\infty}}
\frac{(a^{-1} q^{3/2 + |\tilde{s}|} ; q^2 )_{\infty}}{( aq^{3/2+|\tilde{s}|} ; q ^2)_{\infty}}
\frac{( a^2q ; q^{2} )_{\infty}}{( a^{-2} ; q^{2} )_{\infty}} \; .
\end{align}
By introducing the following function
\begin{equation}
\mathcal{B}(z,m,q^2)=z^{-\frac{1}{4}+\frac{m}{2}}q^{-\frac{1}{8}+\frac{m}{4}}\frac{(zq^{m+1};q^2)_{\infty}}{(z^{-1}q^{m};q^2)_{\infty}} \;,
\end{equation}
one finds a non-trivial pentagon relation
\begin{equation}
\frac{(q^2;q^2)_{\infty}}{(q;q^2)_{\infty}}\oint_{C_0} \,\frac{dz}{2{\pi}iz}z^s\sum^{1}_{m=0}\mathcal{B}(z^{-1}a,m;q^2)\mathcal{B}(za,m;q^2)=\mathcal{B} (a^{-1}q^{-\frac{1}{2}},|\tilde{s}|;q^2)\mathcal{B}(a^{-1}q^{-\frac{1}{2}},|\tilde{s}|+1;q^2)\mathcal{B}(a^2,0;q^2) \; .
\end{equation}

\subsection{Other pentagon identities}

Here we present some other examples of pentagon identities inspired by supersymmetric gauge theory computations. These integral pentagon identities were considered in \cite{Kashaev:2012cz,Gahramanov:2013rda,Gahramanov:2016wxi}.

To obtain a pentagon identity let us consider the following duality. The first theory is the three-dimensional ${\mathcal N}=2$ supersymmetric field theory with $U(1)$ gauge symmetry and $SU(3) \times SU(3)$ flavor group, half of the chirals transforming in the fundamental representation of the gauge group and another half transforming in the anti-fundamental representation. Its dual is a theory with nine chiral multiplets and without gauge degrees of freedom. The supersymmetric duality leads to the following integral identity for hyperbolic hypergeometric functions \cite{Kashaev:2012cz,Spiridonov:2010em,Benvenuti:2016wet}
\begin{equation} \label{indexrel}
 \int^{i\infty}_{-i\infty} dz  \prod_{i=1}^3 s_b(\tfrac{iQ}{2}+a_i+z)  s_b(\tfrac{iQ}{2}+b_i-z)= \prod_{i,j=1}^3s_b(\tfrac{iQ}{2}+a_i+b_j) \; ,
\end{equation}
with the balancing condition $\sum_{i=1}^3(a_i+b_i)=-i Q\,$. Let us introduce the following function
\begin{equation} \label{BB}
\mathcal{B}(x,y) =
\frac{s_b(x+\tfrac{iQ}{2})s_b(y+\tfrac{iQ}{2})}{s_b(x+y+\tfrac{iQ}{2})} \; .
\end{equation}
Then from the expression (\ref{indexrel}) one can easily see that the function  $\mathcal{B}(x,y)$ satisfies the pentagon identity \cite{Kashaev:2012cz}
\begin{equation} \label{Kashaevpent}
\int_{-\textup{i} \infty}^{\textup{i} \infty}dz \prod_{i=1}^3
\mathcal{B}(a_i - z,b_i + z) = \mathcal{B}(a_2+b_1,a_3+b_2)
\mathcal{B}(a_1+b_2,a_3+b_1). 
\end{equation}

We can write similar pentagon relation in terms of basic hypergeometric functions. In order to do that we consider now the $S^2 \times S^1$ partition functions for mirror dual theories. The result is \cite{Gahramanov:2016wxi}
\begin{align} \nonumber
& \sum_{m \in Z}  \oint \frac{dz}{2\pi i z}  (-q)^{\frac12 \sum_{i=1}^3 (\frac{|m_i+m|}{2}+\frac{|n_i-m|}{2}) } z^{- \sum_{i=1}^3 (\frac{|m_i+m|}{2}-\frac{|n_i-m|}{2})}\\ \nonumber
& \qquad \qquad \times  \prod_{i=1}^3  a_i^{-\frac{|m_i+m|}{2}} b_i^{-\frac{|n_i-m|}{2}} \frac{(q^{1+\frac{|m_i+m|}{2}}(a_i z)^{-1};q)_{\infty}}{(q^{\frac{|m_i+m|}{2}} a_i z ;q)_{\infty}} \frac{( q^{1+\frac{|n_i-m|}{2}} z/b_i ;q)_{\infty}}{q^{\frac{|n_i-m|}{2}} (b_i/z ;q)_{\infty}} \\  \label{1int}
& \qquad \qquad  = (-q)^{\frac12 \sum_{i,j=1}^3\frac{|m_i+n_j|}{2}} \prod_{i,j=1}^{3} (a_ib_j)^{-\frac{|m_i+n_j|}{2}} \frac{(q^{1+\frac{|m_i+n_j|}{2}} (a_i b_j)^{-1};q)_{\infty}}{(q^{\frac{|m_i+n_j|}{2}} a_i b_j ;q)_{\infty}}   \;,
\end{align}
with the balancing conditions are 
$\prod_{i=1}^3 a_i  = \prod_{i=1}^3  b_i =q^{\frac12} \;\; \text{and} \;\; \sum_{i=1}^3 n_i  =\sum_{i=1}^3 m_i=0 $.

Again by introducing the following function
\begin{align} \nonumber
{\mathcal B}_m[a, n;b, m] & = (-q)^{\frac{|n|}{4}+\frac{|m|}{4}-\frac{|n+m|}{4}} a^{-\frac{|n|}{2}} b^{-\frac{|m|}{2}} (ab)^{\frac{|n+m|}{2}} \\ 
& \quad \times \frac{(q^{1+\frac{|n|}{2}}a^{-1};q)_\infty}{(q^{\frac{|n|}{2}}a;q)_\infty} \frac{(q^{1+\frac{|m|}{2}}b^{-1};q)_\infty}{(q^{\frac{|m|}{2}}b;q)_\infty} \frac{(q^{\frac{|n+m|}{2}}ab;q)_\infty}{(q^{1+\frac{|n+m|}{2}}(ab)^{-1};q)_\infty} \; ,
\end{align}
one obtains the integral pentagon identity, in terms of ${\mathcal B}$
\begin{align} \nonumber
& \sum_{m\in Z} \oint \frac{d z}{2 \pi i z}   \prod_{i=1}^3 {\mathcal B}[ a_i z, n_i+m; b_i z^{-1}, m_i-m] \\
 & \qquad \qquad = {\mathcal B}[a_1 b_2, n_1+m_2; a_3 b_1; n_3+m_1] \; {\mathcal B}[a_2 b_1, n_2+m_1; a_3 b_2, n_3+m_2] \;.
\end{align}

\section{Concluding remarks}

In this work, we have studied integral pentagon identities which appears in the supersymmetric gauge theory computations. We have focused on examples of supersymmetric dual theories on different manifolds whose partition functions can be written as basic or hyperbolic hypergeometric integrals.

The integral pentagon identity for $\mathbb{RP}^2 \times \mathbb{S}^1$ partition functions are supposed to be related to some invariant of corresponding 3–manifold via $3d-3d$ correspondence that connects three-dimensional $\mathcal N = 2$ supersymmetric theories and triangulated 3–manifolds. It would be interesting to find an interpretation of the pentagon identity in terms of the 3–2 Pachner move for the corresponding 3–manifold in this context.

The identities in terms of hyperbolic hypergeometric functions are related to the knot invariants \cite{Hikami:2006cv,Spiridonov:2011hf} which are connected also to the volumes of hyperbolic 3-manifolds \cite{Kashaev:1996kc} (see also \cite{Gang:2014qla}). It would be interesting to make a connection between knot invariants (as Hikami invariant) and results presented here.

Finally let us mention the intriguing relation of the pentagon identities to integrable models, namely the identities (\ref{indexrel}) and (\ref{1int}) can be written in the form of  the Yang-Baxter equation (see e.g. \cite{Spiridonov:2011hf,Gahramanov:2016ilb,Bazhanov:2016ajm,Jafarzade:2017fsc}). It would be interesting to understand and extend these relations further.

\vspace{0.3cm}

\noindent {\bf Acknowledgments:} Some parts of the paper has been presented at the International Workshop ``Classical and Quantum Integrable Systems'' August 2017. IG thanks the organizers of the workshop (especially V.Spiridonov and P.Pyatov) for the invitation and for creating a stimulating atmosphere. IG would like to thank Hjalmar Rosengren, Vyacheslav Spiridonov and Andrew Kels for useful discussions on the subject of the paper. The authors especially thank to Shahriyar Jafarzade and Dogu D\"{o}nmez for discussions and suggesting valuable improvements on the manuscript. 


\bibliographystyle{utphys}
\bibliography{pentagon,SCI}

\end{document}